# SUPPRESSION OF SPACE CHARGE INDUCED BEAM HALO IN NONLINEAR FOCUSING CHANNEL

Yuri K. Batygin, Alexander Scheinker, Sergey Kurennoy, LANL, Los Alamos, NM 87545, USA

Chao Li, Institute of High Energy Physics, Beijing 100049, China

*Abstract*

An intense non-uniform particle beam exhibits strong emittance growth and halo formation in focusing channels due to nonlinear space charge forces of the beam. This phenomenon limits beam brightness and results in particle losses. The problem is connected with irreversible distortion of phase space volume of the beam in conventional focusing structures due to filamentation in phase space. Emittance growth is accompanied by halo formation in real space, which results in inevitable particle losses. A new approach for solving a self-consistent problem for a matched non-uniform beam in two-dimensional geometry is discussed. The resulting solution is applied to the problem of beam transport, while avoiding emittance growth and halo formation by the use of nonlinear focusing field. Conservation of a beam distribution function is demonstrated analytically and by particle-in-cell simulation for a beam with a realistic beam distribution.

## 1. INTRODUCTION

Traditional accelerator designs utilize linear focusing elements (quadrupoles, solenoids) to yield stable particle motion. However, high intensity rms-matched non-uniform beams are intrinsically mismatched with linear focusing structures. This inconsistency results in space-charge induced beam emittance growth and halo formation. However, periodic structures of focusing-defocusing lenses with combined quadrupole and duodecapole field components provide an effective way to suppress halo formation. This paper summarizes research activity aimed at optimizing the quadrupole-duodecapole channel for halo suppression. The performed analysis allows for matching of a realistic beam with the internal structure of the focusing field. Additionally, beam dynamics studies with a suppressed halo are presented and discussed.

## 2. EFFECT OF BEAM HALO

A beam halo is a comparativley small fraction of the beam (1% - 10%) which lies outside the beam core and simultaneously occupies a significantly larger phase space area than the beam core itself. Accordingly, beam halos constitute a dominant source of beam losses, which result in radio-activation and degradation of accelerator components. Modern accelerator projects that utilize high-intensity beams require the retention of the beam losses at a level of no more than $10^{-7}$/m (less than 1 Watt/m) to avoid activation of the accelerator and to allow hands-on maintenance over long operating periods. Collimation of a beam halo cannot prevent beam losses completely, because the halo of a mismatched beam re-develops in phase space after a certain distance following collimation. Correspondingly, the main sources of beam halo formation in a linac are:

• Mismatch of the beam with the accelerator structure
• Transverse-longitudinal coupling in the RF field
• Misalignments of accelerator channel components
• Aberrations and nonlinearities of focusing elements
• Beam energy tails from un-captured particles
• Particle scattering on residual gas and intra-beam stripping
• Non-linear space-charge forces of the beam.

Halo formation has been a subject of extensive study for several decades (see Refs. [1-9] for a few selected papers). Initial formation of a space-charge induced beam halo is associated with filamentation in phase space due to intrinsic mismatching of a non-uniform beam with its linear focusing structure. Fig. 1 illustrates the injection of a space-charge dominated, non-uniform beam into a focusing channel with a linear quadrupole field, which results in beam uniforming, beam emittance growth, and halo formation ("free energy" effect) using particle-in-cell code BEAMPATH [10]. Halo is formed originally in phase space as long beam tails, and after some oscillations, becomes visible in real space. Further development of halo is associated with particle-core interaction within a focusing channel. In this paper we concentrate on suppression of halo formation in continuous rms-matched, space-charge dominated beams due to charge redistribution within the beam. This issue has been identified as one of the four main reasons in space-charge induced beam emittance growth and halo formation [1]. Beam emittance growth due to "free energy" effect is given by [1]

$$\frac{\varepsilon_f}{\varepsilon_i} = \sqrt{1 + (\frac{\mu_o^2}{\mu^2} - 1)\frac{\Delta W}{W_o}} ,\qquad (1)$$

where $\varepsilon_i$ and $\varepsilon_f$ are initial and final normalized beam emittance, respectively, $\mu_o$ and $\mu$ are the undepressed and space-charge depressed betatron phase advance per period of focusing structure, respectively, and $\Delta W / W_o \sim 0.01....0.077$ is the "free energy " parameter, which depends on the beam distribution. Equation (1) indicates that "free energy" effect is significant for strong space-charge depression factor of $\mu/\mu_o \sim 0.1....0.2$. To exclude other reasons of halo formation, we consider rms-matched, equipartitioned beam, in a perfect focusing structure with equal focusing strengths in *x*- and *y*- directions, far from structure resonances.

### 3. QUADRUPOLE-DUODECAPOLE FOCUSING CHANNEL

Previous studies have shown that effective compensation of the nonlinear space-charge effect can be achieved by adding a duodecapole field component in a quadrupole focusing structure [11]. Such a lattice can be realized by a uniform four-vane structure with specific pole-tip shape imposing duodecapole component in pure quadrupole. The drawback of such a scheme is that duodecapole component can be adjusted only through a change of shape of the electrodes. Here we present a simpler and more practical structure shown in Figure 2. We consider a FODO lattice of the lenses with length *D* and period *L*, with combined quadrupole $G_2(z)$ and duodecapole $G_6(z)$ field components. Such magnets can be realized as a combination of conventional quadrupoles with current sheet magnets [12]. Correspondingly, the physics of particle interaction with focusing structure is different. RFQ-type structure provides continuous focusing, while FODO structure provides discrete focusing with drift spaces between lenses.

As in a previous study, Ref. [11], we treat the problem of suppression of halo formation as a problem of conservation of beam distribution in phase space. The KV beam distribution is the only known *z*-dependent distribution, which is conserved in linear lattice [13]. In Ref. [11] the existence of *z*-independent self-consistent solution for non-KV beam with elliptical symmetry in phase space was demonstrated. In the present study, we extend this approach for analysis of *z*-dependent non-KV beam in a quadrupole-duodecapole FODO channel.

The Hamiltonian of particle motion in a quadruple-duodecapole channel is given by

$$H = \frac{p_x^2 + p_y^2}{2m\gamma} - q\beta c A_z + q\frac{U_b}{\gamma^2} ,\qquad (2)$$

where $p_x$, $p_y$ are components of particle momentum, $U_b$ is the space-charge potential of the beam, $A_z(r,\theta,z)$ is the *z* - component of vector-potential of magnetic field

$$A_z(r,\theta,z) = -[\frac{G_2}{2}r^2\cos(2\theta) + \frac{G_6}{6}r^6\cos(6\theta)]G(z), \tag{3}$$

$G_2$, $G_6$ are gradients of field components of quadrupole and duodecapole components correspondingly:

$$G_m = \frac{B(r_o)}{r_o^{m-1}}, \tag{4}$$

and $G(z)$ is the longitudinal field dependence expanded in Fourier series:

$$G(z) = \frac{4}{\pi}\sum_{k=1}^{\infty}\frac{(-1)^{k-1}}{2k-1}\sin\pi(2k-1)\frac{D}{L}\sin 2\pi(2k-1)\frac{z}{L}. \tag{5}$$

In Cartesian coordinates, the longitudinal component of vector-potential of magnetic field is

$$A_z(x,y,z) = -[\frac{G_2}{2}(x^2 - y^2) + \frac{G_6}{6}(x^6 - y^6 - 15x^4y^2 + 15x^2y^4)]G(z), \tag{6}$$

while components of magnetic field are

$$B_x = \frac{\partial A_z}{\partial y}, \qquad B_y = -\frac{\partial A_z}{\partial x}. \tag{7}$$

The equations of particle motion in this focusing channel are

$$\frac{d^2x}{dt^2} = -\frac{q}{m\gamma}(\beta c B_y + \frac{1}{\gamma^2}\frac{\partial U_b}{\partial x}), \qquad \frac{d^2y}{dt^2} = \frac{q}{m\gamma}(\beta c B_x - \frac{1}{\gamma^2}\frac{\partial U_b}{\partial y}). \tag{8}$$

Taking into account expansion, Eq. (5), the equations of motions can be rewritten as

$$\ddot{\vec{r}} = \frac{q}{m\gamma}[\sum_{k=1}^{\infty}\vec{F}_k(\vec{r})\sin(\omega_k t) - \frac{1}{\gamma^2}\frac{\partial U_b(\vec{r})}{\partial \vec{r}}], \tag{9}$$

where alternative-gradient structure is determined by combination of Fourier harmonics with amplitudes and frequencies

$$\vec{F}_k(\vec{r}) = \frac{(-1)^{k-1}}{2k-1}\frac{4}{\pi}\beta c(-\vec{i}_x B_y + \vec{i}_y B_x)\sin\pi(2k-1)\frac{D}{L}, \tag{10}$$

$$\omega_k = 2\pi(2k-1)\frac{\beta c}{L}. \tag{11}$$

Equation (9) describes the motion in a combination of two oscillating fields, where the first term describes oscillatory field due to variation of quadrupole-duodecapole field, while the second term determines oscillating function due to space-charge potential. However, in channels with betatron phase advance per period, $\mu_o \ll 2\pi$, variation of sizes of matched beam at the period of the structure is small, and function $U_b$ can be approximately assumed to be $z$ - independent. Under these assumptions, according to the averaging method [14], solution of Eq. (9) can be decomposed into a combination of slow variable $\vec{R}(t)$ and a small-amplitude, rapidly oscillating component $\vec{\xi}(t)$:

$$\vec{r}(t) = \vec{R}(t) + \vec{\xi}(t) . \tag{12}$$

In Ref. [13] it is shown that averaging method gives excellent approximation to beam dynamics in lattices with phase advance per focusing period $\mu_o$ up to 60°. The dynamics of the slow variable is governed by the Hamiltonian

$$H = \frac{\dot{\vec{R}}^2}{2} + \frac{q^2}{4(m\gamma)^2} \sum_{k=1}^{\infty} \frac{F_k^2(\vec{R})}{\omega_k^2} + q\frac{U_b}{m\gamma^3} , \tag{13}$$

while the oscillatory correction is determined entirely by the properties of the rapidly varying part of the field:

$$\vec{\xi}(t) = -\frac{q}{m\gamma} \sum_{k=1}^{\infty} \frac{\vec{F}_k(\vec{R})}{\omega_k^2} \sin \omega_k t . \tag{14}$$

Hamiltonian, Eq. (13), can be represented as

$$H = \frac{\dot{\vec{R}}^2}{2} + U_{eff} + q\frac{U_b}{m\gamma^3} , \tag{15}$$

where the effective averaged potential of focusing structure is given by

$$U_{eff} = (\frac{q}{2m\gamma})^2 \sum_{k=1}^{\infty} \frac{F_k^2(\vec{R})}{\omega_k^2} . \tag{16}$$

In the next Section, we derive an explicit expression for effective potential, Eq. (16).

## 4. EFFECTIVE POTENTIAL OF QUADRUPOLE - DUODECAPOLE CHANNEL

Substitution of Eqs. (10), (11), into Eq. (16) gives for effective potential:

$$U_{eff} = (\frac{qL}{\pi^2 m\gamma})^2 (B_x^2 + B_y^2) \sum_{k=1}^{\infty} \frac{1}{(2k-1)^4} \sin^2 \pi(2k-1)\frac{D}{L} . \tag{17}$$

Equation (17) contains square of total magnetic field in the lens

$$B^2 = B_x^2 + B_y^2 = B_r^2 + B_\theta^2 , \tag{18}$$

where radial, $B_r$, and azimuthal, $B_\theta$, components of magnetic field inside lens are expressed through vector potential as

$$B_r = \frac{1}{r}\frac{\partial A_z}{\partial \theta} = G_2 r \sin 2\theta + G_6 r^5 \sin 6\theta , \tag{19}$$

$$B_\theta = -\frac{\partial A_z}{\partial r} = G_2 r \cos 2\theta + G_6 r^5 \cos 6\theta . \tag{20}$$

From Eqs. (19), (20), the square of total magnetic field is

$$B^2 = G_2^2 r^2 + 2G_2 G_6 r^6 \cos 4\theta + G_6^2 r^{10} . \tag{21}$$

To calculate the sum in Eq. (17), we proceed as in Ref. [15]. We expand the sum in Eq. (17) as

$$\sum_{k=1}^{\infty} \frac{\sin^2 \pi (2k-1) \frac{D}{L}}{(2k-1)^4} = \frac{1}{2} [\sum_{k=1}^{\infty} \frac{1}{(2k-1)^4} - \sum_{k=1}^{\infty} \frac{\cos(2k-1)2\pi \frac{D}{L}}{(2k-1)^4}] . \tag{22}$$

The first term in the right hand side of Eq. (22) is [16]

$$\sum_{k=1}^{\infty} \frac{1}{(2k-1)^4} = \frac{\pi^4}{96}, \tag{23}$$

while the second one is [17]

$$\sum_{k=1}^{\infty} \frac{\cos(2k-1)2\pi \frac{D}{L}}{(2k-1)^4} = \frac{\pi^4}{96} - \frac{\pi^4}{4}(\frac{D}{L})^2 + \frac{\pi^4}{3}(\frac{D}{L})^3 . \tag{24}$$

Because the sum, Eq. (24), is given in Ref. [17] without derivations, we present our own prove of correctness of Eq. (24) in Appendix. Combining both Eqs. (23), (24), the expression for sum, Eq. (22), is

$$\sum_{k=1}^{\infty} \frac{\sin^2 (2k-1)\pi \frac{D}{L}}{(2k-1)^4} = \frac{\pi^4}{8}(\frac{D}{L})^2 (1 - \frac{4}{3}\frac{D}{L}) . \tag{25}$$

Finally, the effective potential of the quadrupole-duodecapole focusing structure is:

$$U_{eff}(R,\theta) = \frac{R^2}{2} (\frac{\mu_o \beta c}{L})^2 [1 + 2\eta (\frac{R}{R_b})^4 \cos 4\theta + \eta^2 (\frac{R}{R_b})^8], \tag{26}$$

where $\eta$ is the ratio of field components normalized with respect to two-rms averaged beam size $R_b = 2\sqrt{<x^2>} = 2\sqrt{<y^2>}$:

$$\eta = \frac{G_6}{G_2} R_b^4 , \tag{27}$$

and $\mu_o$ is the averaged phase advance of transverse oscillations in FODO channel [13]:

$$\mu_o = \frac{L}{2D} \sqrt{1 - \frac{4}{3}\frac{D}{L}} \frac{qG_2 D^2}{mc\beta\gamma} . \tag{28}$$

Expression for effective potential, Eq. (26), is similar to that obtained in Ref. [11] for uniform four-vane structure, except for expressions for phase advance $\mu_o$. It indicates that on average the focusing properties of quadruple-duodecapole FODO structure and that of uniform four-vane structure are the same.

## 5. MATCHING OF THE BEAM WITH QUADRUPOLE - DUODECAPOLE CHANNEL

The effective potential has a specific radial and azimuthal dependence (see Fig. 3). In Ref. [11] it was demonstrated that the beam with realistic distribution function, being truncated along equipotential lines of effective potential, Eq. (26), remains approximately matched with the structure. It means that distribution function of such a beam is close to constant and beam propagates without significant emittance growth and halo formation.

Retaining only the leading-order term for the small variable in Eq. (14)

$$\vec{\xi}(t) \approx -\frac{q}{m\gamma}\frac{\vec{F_1}(\vec{R})}{\omega_1^2}\sin\omega_1 t \tag{29}$$

results in the following expressions for components of small-amplitude correction to particle trajectory

$$\frac{\xi_x}{X} = \upsilon_{max}[1+\eta(\frac{X^4 - 10X^2Y^2 + 5Y^4}{R_b^4})]\sin 2\pi\frac{z}{L} , \tag{30}$$

$$\frac{\xi_y}{Y} = -\upsilon_{max}[1+\eta(\frac{Y^4 - 10X^2Y^2 + 5X^4}{R_b^4})]\sin 2\pi\frac{z}{L} , \tag{31}$$

where $X$ and $Y$ are coordinates of slow particle trajectory, and $\upsilon_{max}$ is the relative amplitude of trajectory modulation:

$$\upsilon_{max} = \frac{2}{\pi^2\sqrt{1-\frac{4}{3}\frac{D}{L}}}(\frac{\sin\pi\frac{D}{L}}{\pi\frac{D}{L}})\mu_o . \tag{32}$$

Usage of only the first term in Eq. (14) is justified by numerical simulations presented below. For typical values of $\eta \approx -10^{-2}$, one can neglect nonlinear terms in Eqs. (30), (31). Relative variations of matched beam envelopes in the averaging method are determined by fast oscillating focusing field, and therefore, are independent on beam current and beam emittance [13]. Consequently, relative amplitude of particle oscillations, $\upsilon_{max}$, determines variation of envelopes of matched beam:

$$R_x = R_b(1+\upsilon_{max}\sin 2\pi\frac{z}{L}) , \tag{33}$$

$$R_y = R_b(1-\upsilon_{max}\sin 2\pi\frac{z}{L}) . \tag{34}$$

Consider a beam with parabolic distribution function in four-dimensional phase space, which is qualitatively close to that observed experimentally in the low - energy part of the Los Alamos linear accelerator (see Fig. 4)

$$f = f_o(1-\frac{x^2+y^2}{2R_b^2}-\frac{p_x^2+p_y^2}{2p_o^2}) . \tag{35}$$

Such distribution is characterized by space charge density

$$\rho(r) = \frac{3I}{2\pi\beta c R_b^2}(1 - \frac{r^2}{2R_b^2})^2 , \qquad (36)$$

and phase space density:

$$\rho(x,x') = \frac{3}{2\pi \ni}(1 - \frac{\gamma_x x^2 + 2\alpha_x xx' + \beta_x x'^2}{2 \ni})^2 , \qquad (37)$$

where $\ni = 4 \ni_{rms}$ is the four-rms unnormalized beam emittance and $\alpha_x$, $\beta_x$, $\gamma_x$ are ellipse parameters. As seen in Fig. 4, experimental distribution is characterized by the ratio of total emittance to rms emittance $\ni_{total}/\ni_{rms} = 7.66$, while for parabolic distribution, Eq. (35), this ratio is $\ni_{total}/\ni_{rms} = 8$. In the measurements presented in Fig. 4, a cutoff threshold of 2% of peak value of beam distribution is used to remove experimental noise in the phase space distribution.

In Ref. [18] it was shown that a beam with parabolic distribution is conserved in focusing potential

$$\frac{q}{m\gamma}U = \frac{c^2}{\gamma^2}[\frac{R^2}{2R_b^2}(\frac{\varepsilon^2}{R_b^2} + \frac{3I}{I_c\beta\gamma}) + \frac{3I}{8I_c\beta\gamma}(-\frac{R^4}{R_b^4} + \frac{R^6}{9R_b^6})], \qquad (38)$$

where $I_c = 4\pi\varepsilon_o mc^3/q = 3.13 \times 10^7$ (A/Z) [Amp] is the characteristic beam current, and $\varepsilon = \beta\gamma \ni$ is the four-rms normalized beam emittance. To provide beam matching, quadratic parts of effective potential, Eq. (26), and that of required potential, Eq. (38), have to equal each other, which gives the following expression for phase advance of the focusing structure:

$$\mu_o = \frac{1}{\beta\gamma}\frac{L}{R_b}\sqrt{\frac{\varepsilon^2}{R_b^2} + \frac{3I}{I_c\beta\gamma}} . \qquad (39)$$

The value of duodecapole component is dictated by the condition of equality of both fields at the beam boundary [18]:

$$\eta = -\frac{1}{36(1 + \frac{2}{3b})} , \qquad b = \frac{2}{\beta\gamma}\frac{I}{I_c}(\frac{R_b}{\varepsilon})^2 . \qquad (40)$$

Equations (39), (40), together with equations for beam envelopes, Eqs. (33), (34), and truncation of beam profile along equipotential lines of effective potential, Eq. (26), determine approximate self-consistent solution for conservation of z-dependent beam with parabolic distribution in FODO quadrupole-duodecapole structure. A typical value of the normalized ratio of field components for compensation of nonlinear space charge effect on halo formation is $\eta = -0.015 \ldots -0.027$.

Figure 5 illustrates simulations of a beam in a FODO quadrupole-duodecapole channel using code BEAMPATH [10]. Beam with parabolic distribution, Eq. (35), was truncated along equipotential lines of effective potential, to make the beam distribution close to the matched beam. Presence of periodic duodecapole component of the order $n = 6$ implies selection of phase advance of the lattice below possible resonance value of $360°/6 = 60°$. The channel is characterized by phase advance, Eq. (28), $\mu_o = 21.2°$, space charge tune depression $\mu/\mu_o = 0.1$, and nonlinear field parameter, Eq. (27) $\eta = -0.0156$. In the attached movie [19], CST Particle Studio software simulates 3D beams in structures, presented in Fig. 1 and Fig. 5, with a full solution of Maxwell's equations. For the purposes of this study, a continuous beam was substituted by a finite-length beam portion with an initial length of 1.1×focusing period. Because of the absence of longitudinal focusing, the selected bunch evolves in the longitudinal direction due to longitudinal space charge forces. For that reason, the simulation was restricted to 30 focusing periods. It is clear that the presence of a duodecapole component reduces the amount of halo particles significantly.

For comparison, in Fig. 6 the same beam with truncated distribution was simulated in a pure quadrupole channel. As seen, the beam remains mismatched with a linear focusing field structure. Population of halo is smaller than in case of untruncated beam in a pure quadrupole channel, but noticeably larger than in case of truncated beam in a quadrupole-duodecapole channel (see Fig. 9). This simulation indicates that truncation of beam distribution is not sufficient to suppress beam halo formation, and has to be accompanied with nonlinear focusing elements.

## 6. ADIABATIC MATCHING OF THE BEAM

An additional feature of the proposed structure is its ability to adiabaticaly change the duodecapole component along the channel, which results in a transformation of a truncated, non-uniform beam, into a more uniform beam, matched with the quadrupole focusing structure [18]. Quadrupole field strength is kept constant along the structure while duodecapole component gradually decreases from nominal value to zero at a certain distance. It gives us the possibility to match initially non-unifom beam with the non-linear focusing channel and adiabatically transform it to the beam matched with quadrupole structure. Figs. 7 - 8 illustrate the results of a study of beam parameters as functions of the adiabatic decline distance. Parameters of the beam and the structure were selected to be the same as those in Fig. 5, while duodecapole component dropped linearly from its initial value to zero at a certain distance, measured in numbers of focusing periods, $N_{adiabatic}$. The rest of the structure contained quadrupoles only. Results of the simulations were compared after 100 focusing periods of the structure.

It appears that the optimal starting value of the duodecapole component depends strongly on adiabatic distance. Figure 7 illustrates the dependence of the initial normalized ratio of field components, Eq. (27), as a function of the adiabaticity parameter $N_{adiabatic}$. The dotted line corresponds to the case of constant duodecapole component along the channel. As can be seen, the reduction in the number of adiabatic lengths results in an increase of required duodecapole component.

Figure 8 illustrates particle distributions in phase space and in configuration space after 100 periods of the structure. Figure 8a illustrates halo formation of the beam in a pure quadrupole structure. Figure 8b illustrates the effect of suppression of the halo in quadrupole-duodecapole structure with constant value of the parameter $\eta = -0.0156$ along the channel. Figures 8c – 8d illustrate the effect of adiabatic matching of the beam where the duodecapole component changed adiabatically to zero at the distance of $N_{adiabatic}$ focusing periods. It is clearly visible that the presence of the duodecapole component results in a significant reduction of the beam halo. Figure 9 further illustrates the fraction of particles outside beam core $2.5\sqrt{<x^2>} \cdot 2.5\sqrt{<y^2>}$ for all cases considered in the paper. The fraction of halo particles oscillates along the structure, with significantly reduced number of halo particles in the quadrupole-duodecapole structure compared to a pure quadrupole channel.

## 7. CONCLUSION

In the presented paper we analyze a self-consistent solution for non-KV beam in FODO quadrupole focusing channel with duodecapole component. For proper matching, beam has to be truncated along equipotential lines of effective potential of focusing structure. Conservation of non-uniform beam distribution results in prevention of halo formation and suppression of emittance growth due to "free energy" effect. Such mechanism is significant for beamlines with strong space-charge depression factor of $\mu/\mu_o \sim 0.1....0.2$. Adiabatic decline of the duodecapole component results in transformation of truncated, non-uniform beam, into a beam, which is matched with a pure quadrupole structure. It is found that the required value of duodecapole component strength increases for shorter adiabatic matchers. Further work is required to investigate the robustness of the quadrupole-duodecapole matching method against deviations from the parabolic distribution in phase space.

## 8. ACKNOLEGEMENTS

Authors are indebted to Robert Jameson for valuable discussions and helpful comments.

**APPENDIX**

To calculate the sum, Eq. (24), we use identities:

$$\cos a\vartheta = 1 - a\int_0^\vartheta \sin a\vartheta' d\vartheta', \tag{A1}$$

$$\sin a\vartheta = a\int_0^\vartheta \cos a\vartheta' d\vartheta'. \tag{A2}$$

Combination of Eqs. (A1), (A2) reads

$$\cos a\vartheta = 1 - a^2 \int_0^\vartheta d\vartheta' \int_0^{\vartheta'} d\vartheta'' + a^3 \int_0^\vartheta d\vartheta' \int_0^{\vartheta'} d\vartheta'' \int_0^{\vartheta''} \sin a\vartheta''' d\vartheta'''. \tag{A3}$$

Taking into account that in our notations $\vartheta = \pi D / L$, $a = 2(2k-1)$, we get the following identity:

$$\sum_{k=1}^\infty \frac{\cos 2(2k-1)\vartheta}{(2k-1)^4} = \sum_{k=1}^\infty \frac{1}{(2k-1)^4} - 2\vartheta^2 \sum_{k=1}^\infty \frac{1}{(2k-1)^2} + 8\int_0^\vartheta d\vartheta' \int_0^{\vartheta'} d\vartheta'' \int_0^{\vartheta''} \sum_{k=1}^\infty \frac{\sin 2(2k-1)\vartheta'''}{(2k-1)} d\vartheta'''. \tag{A4}$$

The first term in right hand of Eq. (A4) is given by Eq. (23). The second term reads [16]

$$\sum_{k=1}^\infty \frac{1}{(2k-1)^2} = \frac{\pi^2}{8}. \tag{A5}$$

To evaluate the third term, let us note that it represents Fourier-expansion of periodic step function:

$$\frac{4}{\pi} \sum_{k=1}^\infty \frac{\sin(2k-1)2\vartheta'''}{(2k-1)} = \begin{cases} 1, & 0 \leq 2\vartheta''' \leq \pi, \\ -1, & \pi \leq 2\vartheta''' \leq 2\pi. \end{cases} \tag{A6}$$

Because in FODO focusing channel the length of the lens is limited by that of half of focusing period $0 < D/L \leq 0.5$, the value of $\vartheta''' = \pi D/L$ is limited by $0 < 2\pi D/L \leq \pi$, and in Eq. (A6) the sum is equal to unity. Finally, the sum, Eq. (23) is:

$$\sum_{k=1}^\infty \frac{\cos(2k-1)2\pi\frac{D}{L}}{(2k-1)^4} = \frac{\pi^4}{96} - \frac{\pi^4}{4}(\frac{D}{L})^2 + \frac{\pi^4}{3}(\frac{D}{L})^3. \tag{A7}$$

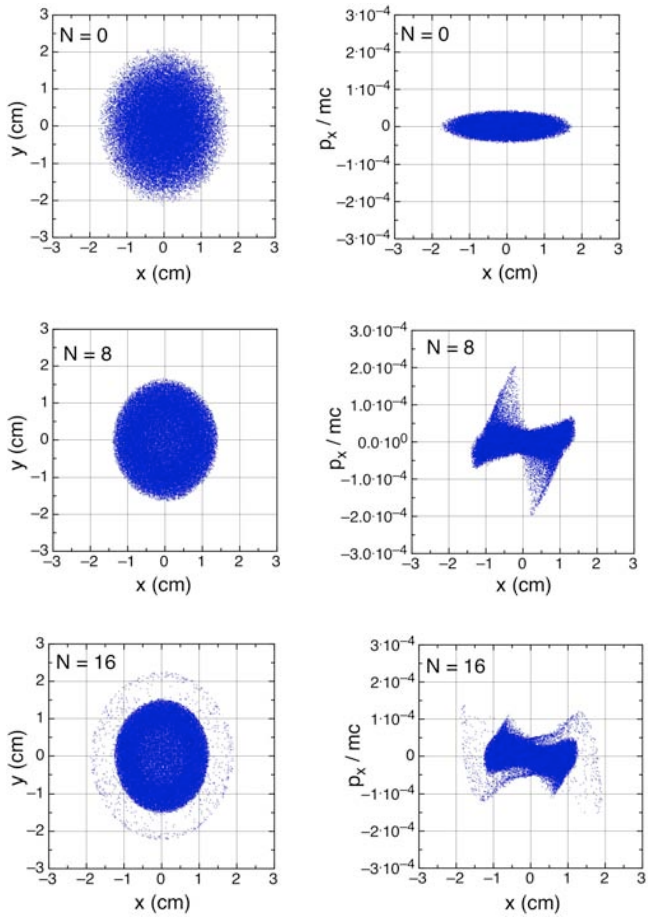

Fig. 1. Emittance growth and halo formation of the beam in FODO quadrupole channel with $\mu_o = 21.2^o$, $\mu/\mu_o = 0.1$: (left) x-y cross section, (right) phase space distribution. Numbers indicate FODO periods.

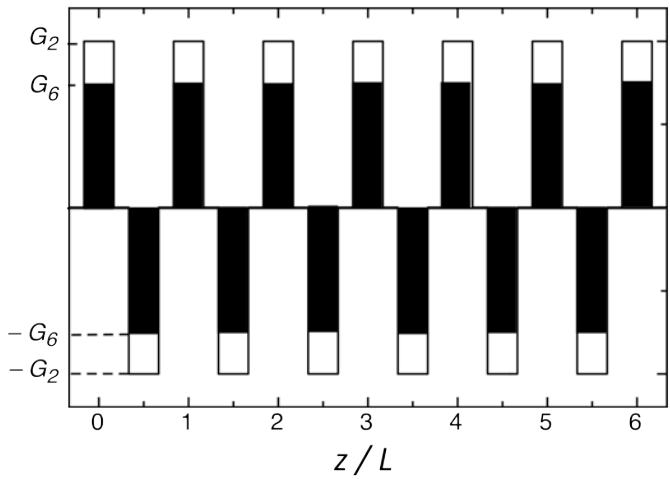

Fig. 2. FODO channel with combined quadrupole $G_2(z)$ and duodecapole $G_6(z)$ field components.

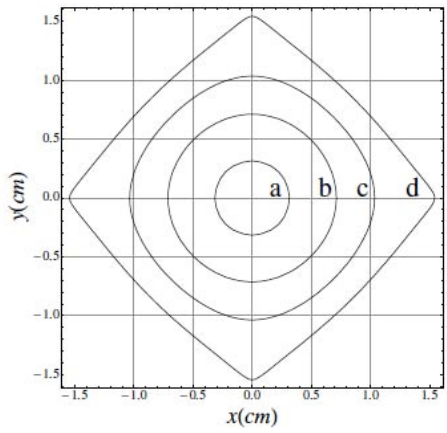

Fig. 3. Lines of equal values of the function $C = \frac{1}{2}r^2 + \zeta r^6 \cos 4\theta + \frac{\zeta^2}{2}r^{10}$ for $\zeta = -0.03$: (a) C = 0.05, (b) C = 0.25, (c) C = 0.5, and (d) C = 0.82.

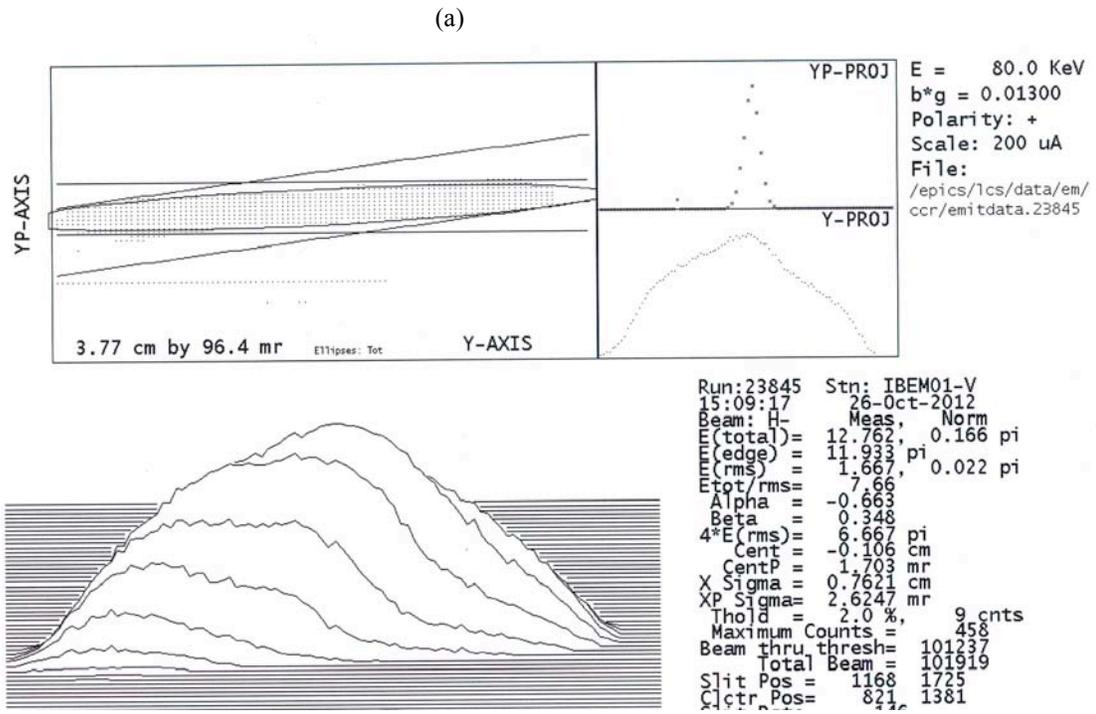

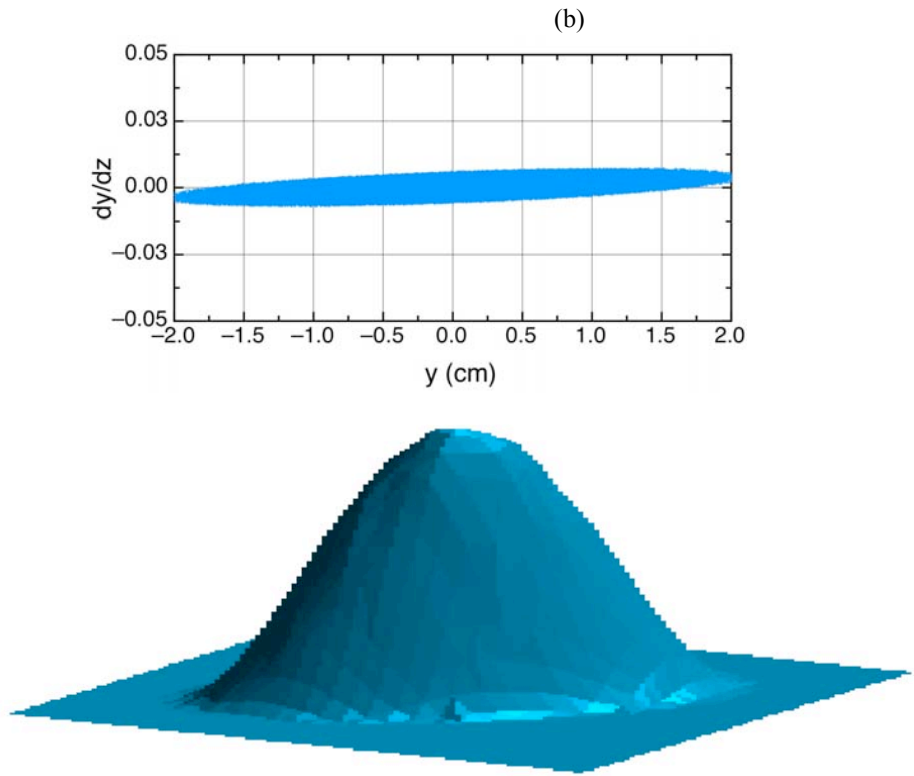

Fig. 4. (a) Experimentally observed distribution of 80 keV H$^-$ beam, extracted from LANL ion source, and (b) modeling of the same beam with parabolic distribution function in 4D phase space, Eq. (35).

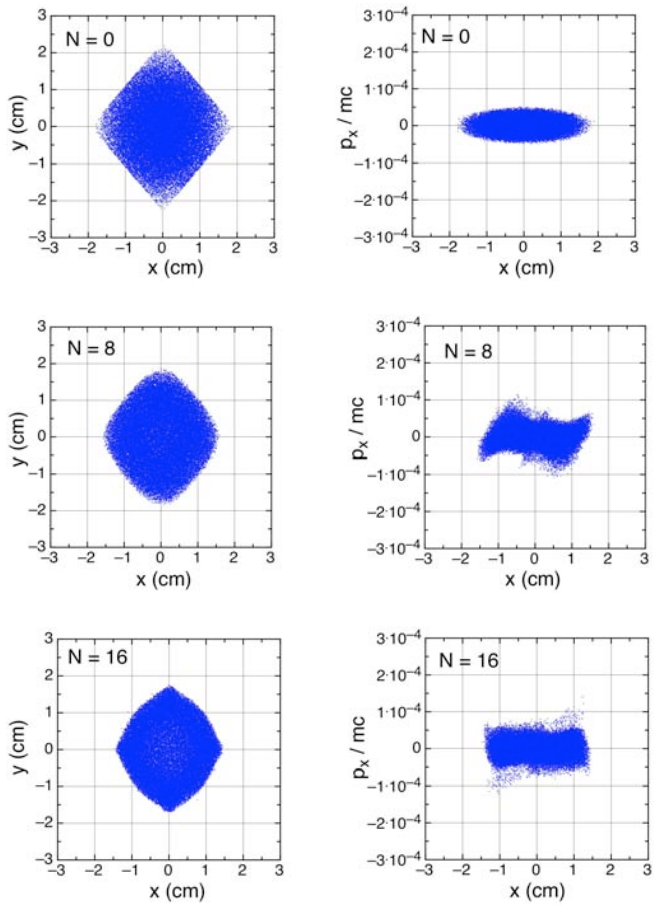

Fig. 5. Suppression of halo formation in FODO quadrupole-dodecapole channel with $\mu_o = 21.2^o$, $\mu/\mu_o = 0.1$, and $\eta = -0.0156$: (left) x-y cross section, (right) phase space distribution. Numbers indicate FODO periods.

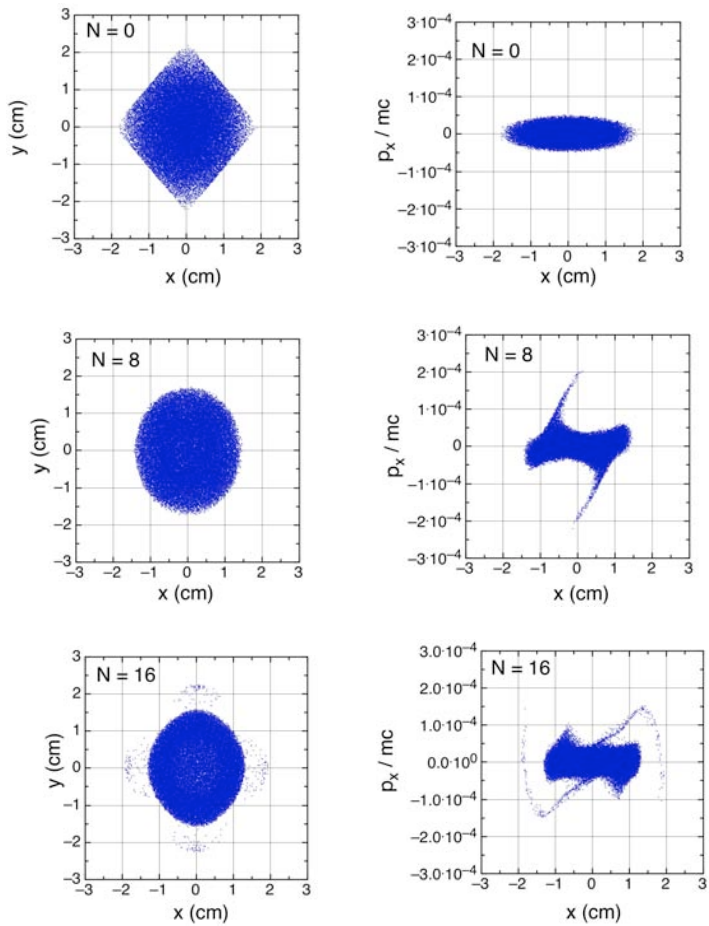

Fig. 6. Emittance growth and halo formation of the truncated beam in FODO quadrupole channel with $\mu_o = 21.2^o$, $\mu/\mu_o = 0.1$: (left) x-y cross section, (right) phase space distribution. Numbers indicate FODO periods.

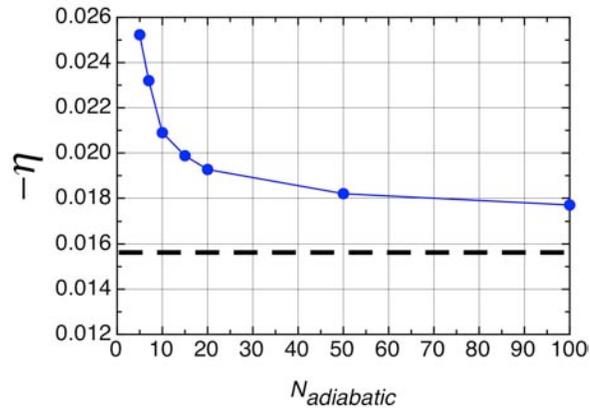

Fig. 7: (solid) optimal starting values of parameter $\eta$, Eq. (27), versus number of FODO periods for adiabatic decline of duodecapole component, (dotted) the value of $\eta$ for the channel with constant duodecapole component.

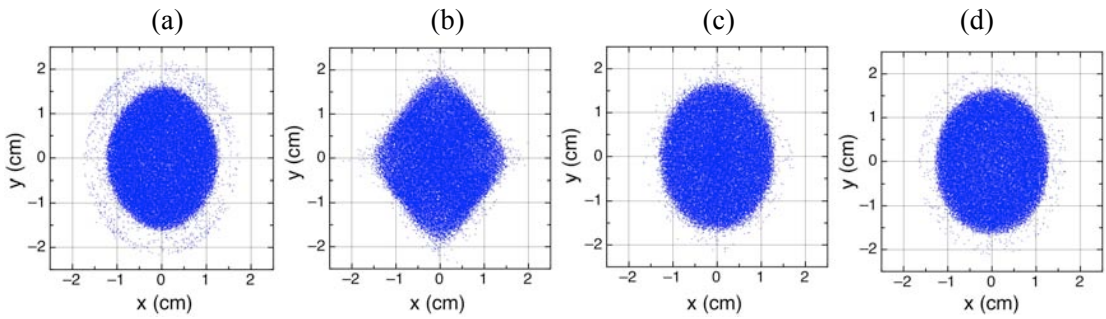

Fig. 8. Particle distributions after 100 focusing periods: (a) pure quadrupole channel, (b) quadrupole-duodecapole channel with constant $\eta = -0.0156$, (c) quadrupole-duodecapole channel with initial $\eta = -0.0177$ and $N_{adiabatic}=100$, (d) quadrupole-duodecapole channel with initial $\eta = -0.0209$ and $N_{adiabatic}=10$.

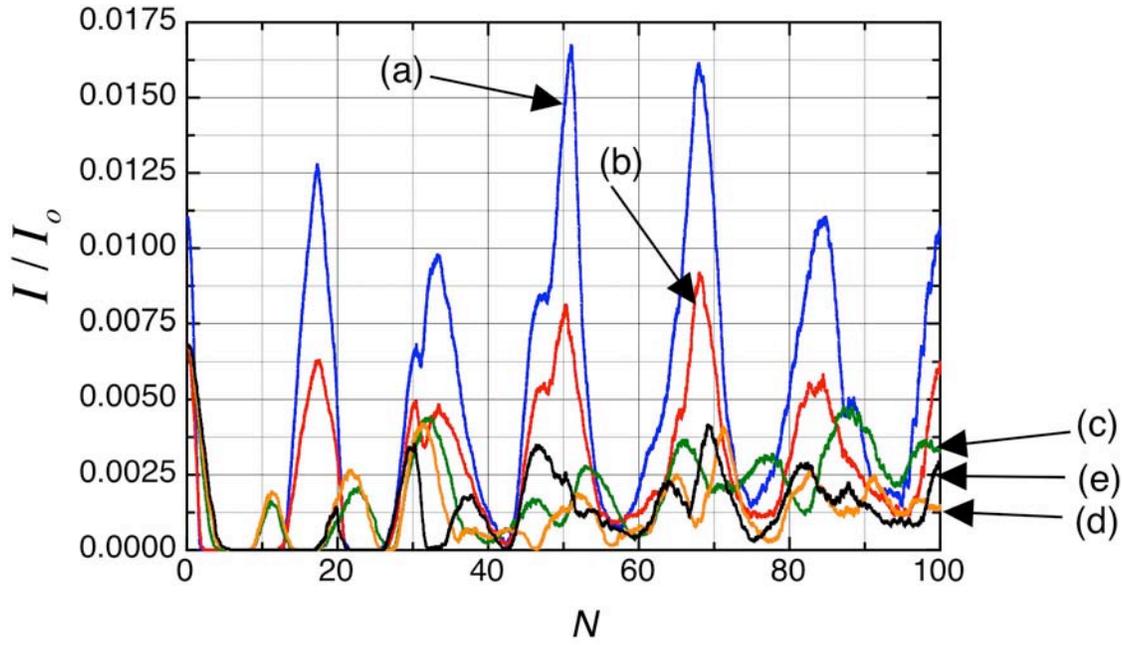

Fig. 9. Fraction of particles outside the beam core $2.5\sqrt{<x^2>}$ x $2.5\sqrt{<y^2>}$ as a function of FODO periods: (a) pure quadrupole channel, see Fig. 1, (b) pure quadrupole channel with truncated beam, see Fig. 6, (c) quadrupole-duodecapole channel with constant $\eta = -0.0156$, see Fig. 5, (d) quadrupole-duodecapole channel with initial $\eta = -0.0177$ and $N_{adiabatic}$=100, see Fig. 8c, (e) quadrupole-duodecapole channel with initial $\eta = -0.0209$ and $N_{adiabatic}$=10, see Fig. 8d.